\documentclass[twocolumn,secnumarabic,amssymb, nobibnotes, aps, prd, showpacs]{revtex4-1}

\usepackage{amsmath}
\usepackage{color}
\usepackage{amssymb}
\usepackage[english]{babel}
\usepackage[latin1]{inputenc}
\usepackage{fancyhdr}
\usepackage{graphics}
\usepackage{fontenc}
\usepackage{graphicx}
\usepackage{makeidx}
\usepackage{ifthen}

\usepackage{bbold}

\usepackage{textcomp}

\usepackage{changes}

\begin{document}

\title{How to calculate the pole expansion of the optical scattering matrix from the resonant states}

\author{T. Weiss}%
\email{t.weiss@pi4.uni-stuttgart.de}
\affiliation{$4^{th}$ Physics Institute and Research Center SCoPE, University of Stuttgart, Pfaffenwaldring 57, D-70550 Stuttgart, Germany}






\author{E. A. Muljarov}%
\affiliation{Cardiff University, School of Physics and Astronomy, The Parade, CF24 3AA, Cardiff, United Kingdom}

\date{\today}%
\begin{abstract}
We present a formulation for the pole expansion of the scattering matrix of open optical resonators, in which the pole contributions are expressed solely in terms of the resonant states, their wavenumbers, and their electromagnetic fields. Particularly, our approach provides an accurate description of the optical scattering matrix without the requirement of a fit for the pole contributions or the restriction to geometries or systems with low Ohmic losses. Hence, it is possible to derive the analytic dependence of the scattering matrix on the wavenumber with low computational effort, which allows for avoiding the artificial frequency discretization of conventional frequency-domain solvers of Maxwell's equations and for finding the optical far- and near-field response based on the physically meaningfull resonant states. This is demonstrated for three test systems, including a chiral arrangement of nanoantennas, for which we calculate the absorption and the circular dichroism.
\end{abstract}

\maketitle


\section{Introduction}

\begin{figure}[htpb]
\begin{center}
\includegraphics[width=\linewidth]{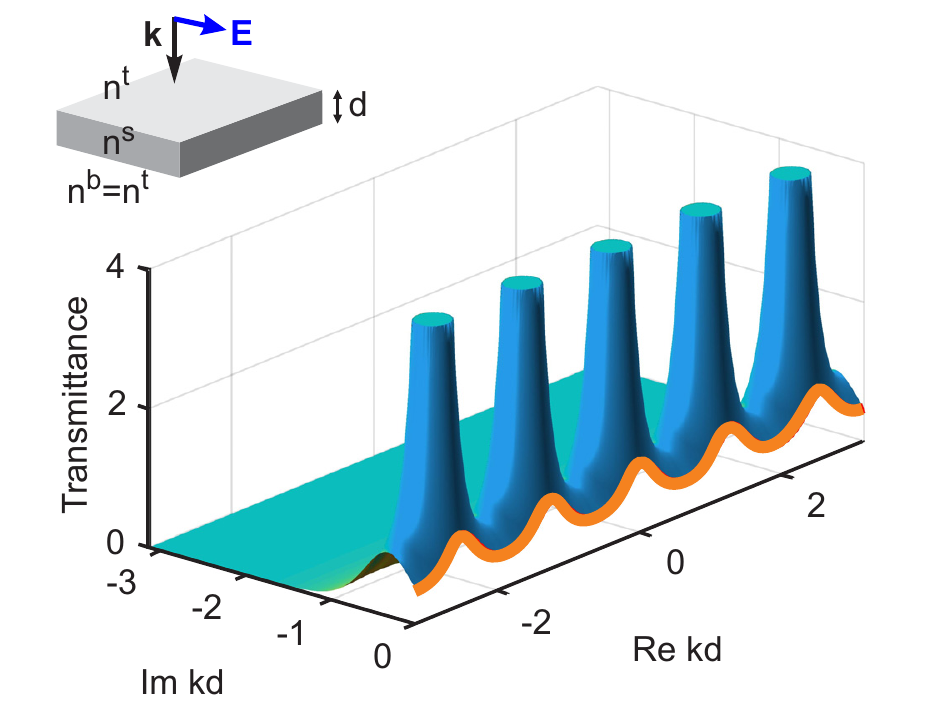}  
\caption{(color online) Analytic continuation of the transmittance for a symmetric planar slab of thickness $d$ with refractive index $n^\mathrm{s}=2.5$ surrounded by homogeneous and isotropic half spaces with index $n^\mathrm{t}=n^\mathrm{b}=1$ at normal incidence (displayed as wavevector $\mathbf{k}$ and electric field $\mathbf{E}$). It can be seen that the oscillating behavior of the transmittance at real wavenumbers $k$ (orange solid line) originates from the poles on the complex $k$ plane, which correspond to the resonant states of that system.} \label{Fig1}
\end{center}
\end{figure}

Since the pioneering work of Gustav Mie~\cite{Mie1908a}, it is known that the scattering of light at small obstacles is governed by the resonant states of that system. Resonant states, also known as quasi-normal modes, are solutions of Maxwell's equations at discrete complex wavenumbers with purely outgoing boundary conditions in the absence of any internal sources. This set of discrete wavenumbers manifests itself as poles in the analytic continuation of any sort of linear optical response function (see Fig.~\ref{Fig1}). According to the Mittag-Leffler theorem~\cite{Arfken2011a}, it is possible to develop a pole expansion for these response functions, including the Green's dyadic as well as the scattering matrix of a system. Knowing the poles and residues of the response function then provides its behavior as a function of wavenumber, which is clearly advantageous compared to conventional numerical calculations, where an artificial discretization in either time- or frequency-domain is introduced~\cite{Moharam1981a,Draine1988a,Hafner1995a,Martin1998a,Whittaker1999a,Tikhodeev2002a,Niegemann2009a}.

Regarding the Green's dyadic, it has been shown that the residues of its pole expansion can be derived from the resonant electric field distributions of the resonant states when normalizing them appropriately~\cite{Muljarov2010a}. Several approaches have been suggested for the normalization of resonant states~\cite{Sammut1976a, Lai1990a, Weinstein1969a, Sauvan2013a, Muljarov2010a, Doost2013a, Armitage2014a, Doost2014a, Muljarov2016a, Muljarov2016b, Weiss2016a,Weiss2017a,Bai2013a, Kristensen2012a, Kristensen2015a,Muljarov2017a,Muljarov2018a}. This includes approximate formulations for high-quality modes~\cite{Kristensen2012a, Kristensen2015a,Muljarov2017a}, the utilization of perfectly matched layers~\cite{Sauvan2013a} or, equivalently, complex coordinates~\cite{Sammut1976a} in the exterior of the system, as well as numerical approaches~\cite{Bai2013a,Weiss2016a}. A fully analytical form of normalization has been derived in~\cite{Muljarov2010a} and extended to various geometries and materials~\cite{Doost2012a,Doost2013a, Armitage2014a, Doost2014a,Muljarov2016a,Muljarov2016b,Weiss2016a,Weiss2017a,Lobanov2017a,Muljarov2018a}.

Existing methods for expanding the scattering matrix in terms of its poles often require fitting procedures or corrections based on symmetry arguments~\cite{Fan2003a,Gippius2005a}, and have been proven only for single resonant states~\cite{Fan2003a,Gippius2005a,Ruan2012a}. The approach of Perrin~\cite{Perrin2016a} for the pole expansion of a scattered field is more general and has been validated for two poles. In a similar manner, Yang et al. show how to derive the outgoing channels for a known electromagnetic near field~\cite{Yang2015a}, but both Perrin's and Yang's approaches do not provide the pole expansion of the scattering matrix, which directly relates incoming and outgoing channels. Recently, Alpeggiani et al.~\cite{Alpeggiani2017a} presented a formulation  that is valid for a large number of resonant states. Based on symmetry considerations, a system of equations is derived that is solved in a least-square sense in order to calculate the residues for the pole expansion of the scattering matrix. However, the approach cannot be applied to strongly absorbing systems, with the least-square method being an artificial fitting procedure that increases the computational time proportional to the number of considered resonant states and channels in the scattering matrix.

Here, we derive the pole expansion of the scattering matrix from the pole expansion of the Green's dyadic for reciprocal systems. In contrast to previous works, the residues of the pole contributions in the scattering matrix are calculated directly from the resonant field distributions by projecting them onto basis functions of free space~\cite{Yang2015b}. The derivation is based on selecting basis functions that satisfy appropriate orthogonality relations, and separating the total field into the background and scattered field~\cite{Martin1998a,Rosenkrantz2013a,Perrin2016a,Alpeggiani2017a}. The formulation is neither limited with respect to the number of resonant states nor restricted to certain geometries. Moreover, it is valid for strongly absorbing systems and can be implemented in any numerical frequency-domain solver for Maxwell's equations that allows for calculating the resonant states and their field distributions. Thus, it is possible to derive the scattering matrix for  arbitrary geometries as a function of wavenumber, which is particularly useful for systems that exhibit resonant states with a narrow linewidth such as Fano resonances~\cite{Fano1961a,Liu2009a,Gallinet2011a}. In addition, our approach can be applied to systems with resonant phenomena dominated by loss, such as perfect absorbers~\cite{Liu2010b} and chiral nanoantenna arrays~\cite{Yin2013a,Fernandez2016a}.

The paper is organized as follows: In Section~\ref{GDsec}, we give an overview of the compact operator form of Maxwell's equations introduced in~\cite{Muljarov2018a} and define two forms of bilinear maps in order to simplify the further derivation steps. Section~\ref{RS} is devoted to the resonant states and the pole expansion of the Green's dyadic. After this introduction, we recapitulate in Section~\ref{BGsec} the concept of background field and scattered field. In Section~\ref{Ssec}, we define the scattering matrix as well as the orthogonality relations of the basis functions selected as input and output channels of the scattering matrix. Using the notations and relations of Sections~\ref{GDsec} to~\ref{Ssec}, we then derive in Section~\ref{Psec} the pole expansion of the scattering matrix for arbitrary geometries, which is the central result of this work. Particularly, we show that it is possible to calculate the residues of the pole expansion solely from the resonant field distributions. In Sections~\ref{PSsec} and~\ref{Rsec}, we focus on planar periodic systems, for which we compare the results of our formulation with full numerical and analytical calculations for three test systems: A planar symmetric slab consisting of homogeneous and isotropic materials, a dielectric grating with quasi-guided modes~\cite{Gippius2010a}, and a chiral arrangement of gold wire antennas~\cite{Yin2013a}. The full numerical calculations are based on the Fourier modal method with adaptive coordinates~\cite{Granet1999a,Granet2002a,Weiss2009a,Weiss2009b,Essig2010a}. The resonant states and their field distributions are calculated by the methods described in~\cite{Weiss2011a,Bykov2012a,Weiss2016a}. In the Appendix, we sketch how to use our formalism for single scatterers in three-dimensional space and give more details on the planar periodic systems.


\section{Maxwell's equations and Green's dyadic}\label{GDsec}

In Ref.~\cite{Muljarov2018a}, a compact matrix-operator formulation of Maxwell's equations has been introduced, which is given in Gaussian units and frequency domain [time dependence $\exp(-i\omega t)$] by
\begin{equation}
\hat{\mathbb{M}}(\mathbf{r};k) \mathbb{F}(\mathbf{r};k) =  \mathbb{J}(\mathbf{r};k), \label{Maxwell}
\end{equation}
where $k=\omega/c$ is the wavenumber and $\hat{\mathbb{M}}(\mathbf{r};k)=k\hat{\mathbb{P}}(\mathbf{r};k)-\hat{\mathbb{D}}(\mathbf{r})$ with
\begin{align}
\hat{\mathbb{P}}(\mathbf{r};k) &= \left[\begin{array}{cc} \varepsilon(\mathbf{r};k) & -i\xi(\mathbf{r};k) \\ i\zeta(\mathbf{r};k) & \mu(\mathbf{r};k)\end{array}\right], \\ \hat{\mathbb{D}}(\mathbf{r}) &= \left(\begin{array}{cc} 0 & \nabla\times \\ \nabla\times & 0 \end{array}\right).
\end{align}
In general, $\varepsilon$, $\mu$, $\zeta$, and $\xi$ are $3\times3$ tensors. For reciprocal materials, the bi-anisotropy tensors $\xi$ and $\zeta$ obey $\xi^\mathrm{T}=-\zeta$, whereas $\varepsilon^\mathrm{T}=\varepsilon$ and $\mu^\mathrm{T}=\mu$, so that $\hat{\mathbb{P}}^\mathrm{T} = \hat{\mathbb{P}}$, with the superscript T denoting the matrix transpose. The electric and magnetic fields as well as the currents form six-dimensional supervectors:
\begin{align}
\mathbb{F}(\mathbf{r};k) &= \left[\begin{array}{c} \mathbf{E}(\mathbf{r};k) \\ i\mathbf{H}(\mathbf{r};k) \end{array}\right], & \mathbb{J}(\mathbf{r};k) &= \left[\begin{array}{c} \mathbf{J}_\mathrm{E}(\mathbf{r};k) \\ i\mathbf{J}_\mathrm{H}(\mathbf{r};k) \end{array}\right].
\end{align}
Here, $\mathbf{J}_\mathrm{E}(\mathbf{r};k)=-4\pi i \mathbf{j}(\mathrm{r};k) / c$, and the magnetic currents $\mathbf{J}_\mathrm{H}$ have been introduced for symmetry purposes.

Equation~(\ref{Maxwell}) presents an inhomogeneous linear differential equation. Its Green's dyadic $\hat{\mathbb{G}}(\mathbf{r},\mathbf{r}';k)$ fulfills
\begin{equation}
\hat{\mathbb{M}}(\mathbf{r};k) \hat{\mathbb{G}}(\mathbf{r},\mathbf{r}';k) = \mathbb{1}\delta(\mathbf{r}-\mathbf{r}'),
\end{equation}
which allows to construct a special solution of Eq.~(\ref{Maxwell}) as
\begin{equation}
\mathbb{F}(\mathbf{r};k) = \int\limits_V\mathrm{d}V' \hat{\mathbb{G}} (\mathbf{r},\mathbf{r}';k) \mathbb{J}(\mathbf{r}';k). \label{GreenSolv}
\end{equation}

For later convenience, we define two types of bilinear maps between two six-dimensional field supervectors $\mathbb{A}$ and $\mathbb{B}$ that are either field vectors $\mathbb{F}$ or current vectors $\mathbb{J}$:
\begin{align}
\mathbb{A} &= \left(\begin{array}{c} \mathbf{A}_\mathrm{E} \\ i\mathbf{A}_\mathrm{H} \end{array}\right), & \mathbb{B} &= \left(\begin{array}{c} \mathbf{B}_\mathrm{E} \\ i\mathbf{B}_\mathrm{H}\end{array}\right).
\end{align} 
The first bilinear map is defined as a volume integral over a finite volume $V$:
\begin{equation}
\langle \mathbb{A} | \mathbb{B}\rangle_V \equiv \int\limits_V \mathrm{d}V \left(\mathbf{A}_\mathrm{E}\cdot\mathbf{B}_\mathrm{E} - \mathbf{A}_\mathrm{H}\cdot\mathbf{B}_\mathrm{H} \right). \label{Bilinear1}
\end{equation}
This bilinear map is symmetric, i.e., $\langle \mathbb{A} | \mathbb{B}\rangle_V =\langle \mathbb{B} | \mathbb{A}\rangle_V $. The second one is defined as a surface integral on the surrounding surface $\partial V$ of $V$:
\begin{equation}
[\mathbb{A}|\mathbb{B}]_{\partial \mathcal{V}} \equiv i\oint\limits_{\partial \mathcal{V}} \!\mathrm{d} \mathbf{S} \cdot  \left(\mathbf{A}_\mathrm{E}\!\times\!\mathbf{B}_\mathrm{H} \!-\! \mathbf{B}_\mathrm{E}\!\times\!\mathbf{A}_\mathrm{H} \right) . \label{Bilinear2}
\end{equation}
It is antisymmetric with $[\mathbb{A}|\mathbb{B}]_{\partial \mathcal{V}} = -[\mathbb{B}|\mathbb{A}]_{\partial \mathcal{V}}$, and equals zero for identical vectors, i.e., $[\mathbb{A}|\mathbb{A}]_{\partial \mathcal{V}} =0$. Using vector identities and the divergence theorem, it is straight-forward to show that
\begin{equation}
[\mathbb{A}|\mathbb{B}]_{\partial \mathcal{V}} =  \langle \mathbb{B} |\hat{\mathbb{D}}|\mathbb{A}\rangle_V -\langle \mathbb{A} |\hat{\mathbb{D}}|\mathbb{B}\rangle_V. \label{Bilinear3}
\end{equation}

\section{Resonant states}\label{RS}

\begin{figure}[htpb]
\begin{center}
\includegraphics[width=\linewidth]{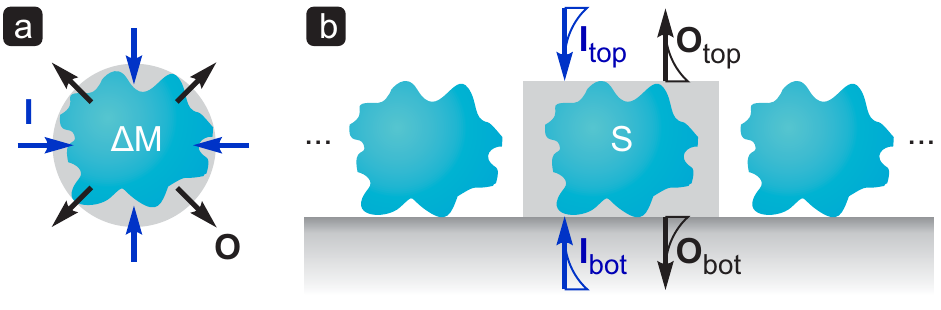}  
\caption{(color online) Typical scattering geometries with spatial inhomogeneities $\Delta\hat{\mathbb{M}}$ denoted by light blue color: (a) Single scatterer in homogeneous and isotropic space, (b) planar periodic system with super- and substrate at the top and bottom, respectively. The scattering matrix $\mathcal{S}$ provides a connection between the incoming fields $\vec{\mathcal{I}}$ and the outgoing fields $\vec{\mathcal{O}}$. The residues for the pole expansion of the scattering matrix should be calculated from the resonant field distributions on the surface of a minimal convex volume surrounding the scatterer, which is denoted by the light gray regions.} \label{Fig2}
\end{center}
\end{figure}

The homogeneous form of Eq.~(\ref{Maxwell}) possesses a countable number of solutions on the complex $k$ plane satisfying outgoing boundary conditions, which are the resonant states. They obey the equation
\begin{equation}
\hat{\mathbb{M}}(\mathbf{r};k_n) \mathbb{F}_n(\mathbf{r}) = 0. \label{ResDef}
\end{equation}
The real part of $k_n$ gives the resonance wavenumber, whereas $-2\mathrm{Im}k_n$ is the resonance linewidth. When normalizing the resonant states appropriately~\cite{Muljarov2010a,Doost2012a,Doost2013a,Armitage2014a, Doost2014a,Muljarov2016a,Muljarov2016b,Weiss2016a,Weiss2017a,Muljarov2018a}, they can be used together with possible cuts in order to expand the Greens dyadic with outgoing boundary conditions as follows:
\begin{equation}
\hat{\mathbb{G}}(\mathbf{r},\mathbf{r}';k) = \sum\limits_n \frac{\mathbb{F}_n(\mathbf{r}) \otimes \mathbb{F}_n^\mathrm{R}(\mathbf{r}')}{k-k_n} . \label{GreenExp}
\end{equation}
Here, the superscript $\mathrm{R}$ denotes the reciprocal conjugate resonant states that are solutions of Maxwell's equations at the same wavenumber $k_n$ but for reciprocal boundary conditions (which corresponds to a reversal in pathways such as in-plane momentum $\mathbf{k}_{||}\rightarrow-\mathbf{k}_{||}$ for planar systems~\cite{Weiss2017a}), and $\otimes$ is the outer vector product. Note that we do not distinguish explicitly between pole and cut contributions, since they can be formally written in the same way and replaced by a finite number of cut poles in numerical calculations~\cite{Doost2013a,Lobanov2017a}. Furthermore, it has to be emphasized that the pole expansion of the Green's dyadic in Eq.~(\ref{GreenExp}) is generally not valid in the region outside the scatterer, since the resonant states contributing to Eq.~(\ref{GreenExp}) contain only outgoing waves, as discussed in more detail in~\cite{Weiss2017a}. In order to warrant its applicability, it should be restricted to a minimal convex volume enclosing the scatterer, see Fig.~\ref{Fig2}. 

The most general form of the analytical normalization condition can be written as~\cite{Muljarov2018a}
\begin{equation}
1=\langle \mathbb{F}_n^\mathrm{R} |( k\hat{\mathbb{P}})' |\mathbb{F}_n\rangle_{V_\mathrm{N}} + [\mathbb{F}_n^\mathrm{R}|\mathbb{F}'_n]_{\partial V_\mathrm{N}} , \label{Norm}
\end{equation}
where $V_\mathrm{N}$ is the volume of normalization, and the prime denotes the derivative with respect to $k$ at $k_n$. Calculating the derivative is trivial for the first term on the right hand side. For the second term, we have to differentiate the analytical continuation of $\mathbb{F}_n$ on the complex $k$~plane, which depends on the geometry of interest~\cite{Doost2012a,Doost2013a,Armitage2014a, Doost2014a,Weiss2016a,Weiss2017a}.

It should be mentioned that we are using here a slightly different formulation~\cite{Muljarov2018a} of the normalization than in most of our previous works on the resonant state expansion~\cite{Muljarov2010a, Doost2012a, Doost2013a, Armitage2014a, Doost2014a, Muljarov2016a, Muljarov2016b, Weiss2016a,Weiss2017a,Muljarov2017a,Lobanov2017a}, which is valid also for magnetic and bi-anisotropic materials. This formulation can be reduced to our previous results for nonmagnetic materials that are solely described by the electric field, the electric permittivity, and the electric current as a special case. Note, however, that in the new formulation, the normalized electric field is a factor of $\sqrt{2}$ smaller than in previous works.

\section{Background and scattered field}\label{BGsec}

For a given system, Maxwell's operator $\hat{\mathbb{M}}$ in Eq.~(\ref{Maxwell}) can be separated into a background term $\hat{\mathbb{M}}_\mathrm{BG}=k\hat{\mathbb{P}}_\mathrm{BG}-\hat{\mathbb{D}}$ for a simple material distribution, and a scatterer $\Delta\hat{\mathbb{M}}=k(\hat{\mathbb{P}}-\hat{\mathbb{P}}_\mathrm{BG})$, which describes a spatial inhomogeneity (see Fig.~\ref{Fig2}). The field scattered at $\Delta\hat{\mathbb{M}}$ can be obtained by separating the total field into a background field $\mathbb{F}_\mathrm{BG}$ that is a solution of the background Maxwell's equations for a given incoming field,
\begin{equation}
\hat{\mathbb{M}}_\mathrm{BG}(\mathbf{r};k)\mathbb{F}_\mathrm{BG}(\mathbf{r};k)=0, \label{Background}
\end{equation}
and the scattered field  $\mathbb{F}_\mathrm{scat}$~\cite{Martin1998a,Rosenkrantz2013a,Perrin2016a,Alpeggiani2017a}:
\begin{equation}
\mathbb{F}(\mathbf{r};k) = \mathbb{F}_\mathrm{BG}(\mathbf{r};k) + \mathbb{F}_\mathrm{scat}(\mathbf{r};k). \label{TotalField}
\end{equation}

Using Eq.~(\ref{Background}) in Maxwell's equations for the full system, it is straight-forward to obtain
\begin{equation}
\hat{\mathbb{M}}(\mathbf{r};k)\mathbb{F}_\mathrm{scat}(\mathbf{r};k) = -\Delta\hat{\mathbb{M}}(\mathbf{r};k)\mathbb{F}_\mathrm{BG}(\mathbf{r};k). \label{GetF}
\end{equation}
Thus, knowing the pole expansion Eq.~(\ref{GreenExp}) for the Green's dyadic of $\hat{\mathbb{M}}$ provides via Eq.~(\ref{GreenSolv}) the scattered field within the system as~\cite{Perrin2016a}
\begin{equation}
\mathbb{F}_\mathrm{scat}(\mathbf{r};k) = -\sum\limits_n \mathbb{F}_n(\mathbf{r})\frac{\langle \mathbb{F}_n^\mathrm{R}|\Delta\hat{\mathbb{M}}|\mathbb{F}_\mathrm{BG}\rangle_V}{k-k_n} .\label{Fexp}
\end{equation}
It should be mentioned that the background field $\mathbb{F}_\mathrm{BG}$ possesses a regular behavior inside the scatterer, i.e., it does not diverge at any point, whereas the corresponding incoming and outgoing fields may have singularities such as the incoming and outgoing Hankel functions. For more details, see Appendix~\ref{Spherical}.

\section{The scattering matrix}\label{Ssec}

The scattering matrix provides a relation between incoming and outgoing channels of a system. A channel is defined by a solution of Maxwell's equations in the surrounding of the scatterer, which we assume to be homogeneous and isotropic space or two half spaces of homogeneous and isotropic dielectric materials separated by a planar interface (see Fig.~\ref{Fig2}). For such cases, Maxwell's equations provide complete sets of orthogonal basis functions that can be used to expand an arbitrary solution of Maxwell's equations outside the system and on the surface surrounding the scatterer. The best-suited basis set depends on the geometry: Plane waves for planar systems, vector spherical harmonics for finite three-dimensional geometries, and cylindrical waves for systems with translational symmetry in one dimension.

Each basis function defines one incoming and one outgoing channel. From a mathematical point of view, the definition of incoming and outgoing channels is arbitrary, and more motivated by physical arguments. For real-valued wavenumbers $k$, an outgoing channel either possesses a time-averaged energy flux that propagates outwards, or the fields are decaying exponentially with distance to the scatterer. The analytical continuation to the complex $k$ plane is more sophisticated~\cite{Akimov2011a,Weiss2017a}, but usually defined such that there is a smooth transition from any point on the complex $k$ plane to the real axis. Thus, an arbitrary field in the exterior can be expanded as follows:
\begin{equation}
\mathbb{F}(\mathbf{r};k) = \sum\limits_{\mathbf{N}} \alpha_{\mathbf{N}}^\mathrm{in}(k) \mathbb{I}_{\mathbf{N}}(\mathbf{r};k) + \alpha_{\mathbf{N}}^\mathrm{out}(k) \mathbb{O}_{\mathbf{N}}(\mathbf{r};k). \label{ArbFieldExp}
\end{equation}
The index $\mathbf{N}$ denotes a set of quantum numbers that specify the different basis functions $\mathbb{I}_{\mathbf{N}}$ and $\mathbb{O}_{\mathbf{N}}$, respectively, of the incoming and outgoing nature. The scattering matrix $\mathcal{S}$ then relates incoming and outgoing channels as follows:
\begin{equation}
\vec{\mathcal{O}}(k) = \mathcal{S}(k) \vec{\mathcal{I}}(k). \label{Smatrix}
\end{equation}
Here, $\vec{\mathcal{I}}$ and $\vec{\mathcal{O}}$ are supervectors containing the expansion coefficients $\alpha_{\mathbf{N}}^\mathrm{in}$ and $\alpha_{\mathbf{N}}^\mathrm{out}$, respectively. The elements of the scattering matrix are labeled as $\mathcal{S}_{\mathbf{N}\mathbf{N}'}$ for an incoming channel with quantum number $\mathbf{N}'$ and an outgoing channel with quantum number $\mathbf{N}$.

How can we derive the expansion coefficients $\alpha_{\mathbf{N}}^\mathrm{in}$ and $\alpha_{\mathbf{N}}^\mathrm{out}$? This can be achieved by introducing, following Ref.~\cite{Yang2015b}, a set of orthogonal modes in free space. Based on the reciprocity principle, there exist reciprocal conjugate basis functions $\mathbb{I}_{\mathbf{N}}^\mathrm{R}$ and $\mathbb{O}_{\mathbf{N}}^\mathrm{R}$ such that
\begin{align}
[\mathbb{I}_{\mathbf{N}}^\mathrm{R} | \mathbb{O}_{\mathbf{N}'}]_{\partial V} = -[\mathbb{O}_{\mathbf{N}}^\mathrm{R} | \mathbb{I}_{\mathbf{N}'}]_{\partial V} &= \delta_{\mathbf{N},\mathbf{N}'} , \label{Orth1} \\
[\mathbb{I}_{\mathbf{N}}^\mathrm{R} | \mathbb{I}_{\mathbf{N}'}]_{\partial V} = [\mathbb{O}_{\mathbf{N}}^\mathrm{R} | \mathbb{O}_{\mathbf{N}'}]_{\partial V} &= 0 .\label{Orth2}
\end{align}
At a first glance, it might seem strange that we have to combine incoming and outgoing basis functions in order to obtain unity for $\mathbf{N}'=\mathbf{N}$. However, this is a direct consequence of the antisymmetric form of Eq.~(\ref{Bilinear2}). From the physical point of view, it is related to the fact that reciprocity provides the connection between fields of reciprocal pathways. In general, the orthogonality given by Eqs.~(\ref{Orth1}) and~(\ref{Orth2}) follows from the fact that the proper choice of the basis functions in free space determining its quantum numbers $\mathbf{N}$ is dictated by the properties of the surface $\partial V$ chosen for the integration, and, in particular, by its symmetry. While we do not provide a general proof for an arbitrary surface, we prove in Appendices~\ref{Spherical} and~\ref{OrthPW} the orthonormality Eqs.~(\ref{Orth1}) and~(\ref{Orth2}) in two important cases: A spherical surface $\partial V$ and the surface $\partial V$ of a unit cell of a planar periodic system. For $\mathbf{N}=\mathbf{N}'$, equation~(\ref{Orth1}) determines the normalization of modes in free space. Using Eqs.~(\ref{Orth1}) and~(\ref{Orth2}) in Eq.~(\ref{ArbFieldExp}), we obtain: 
\begin{align}
\alpha_{\mathbf{N}}^\mathrm{in}(k) &= [\mathbb{O}_{\mathbf{N}}^\mathrm{R} | \mathbb{F}]_{\partial V}, & \alpha_{\mathbf{N}}^\mathrm{out}(k) &= [\mathbb{I}_{\mathbf{N}}^\mathrm{R} | \mathbb{F}]_{\partial V}. \label{ExpCoeff}
\end{align}

\section{Pole expansion}\label{Psec}

Formally, the pole expansion of the scattering matrix yields: 
\begin{equation}
\mathcal{S}(k) = \mathcal{S}_\mathrm{BG} + \sum\limits_n \frac{\mathcal{R}_n}{k-k_n} . \label{Sexp}
\end{equation}
The background term $\mathcal{S}_\mathrm{BG}$ arises, because  even for the vacuum background, its scattering matrix is nonzero. As in~\cite{Alpeggiani2017a}, where $\mathcal{S}_\mathrm{BG}$ is called the direct-coupling matrix, we do not focus on the calculation of the background term, but derive expressions for the residue matrices $\mathcal{R}_n$ with elements $\mathcal{R}_{n,\mathbf{N}\mathbf{N}'}$.

Applying Eq.~(\ref{ExpCoeff}) to Eq.~(\ref{TotalField}), it is possible to calculate, for a given incident field, the expansion coefficients of the outgoing field for a scattering geometry as the elements of the vector $\vec{\mathcal{O}}$ in Eq.~(\ref{Smatrix}):
\begin{equation}
\alpha_{\mathbf{N}}^\mathrm{out}(k) = [\mathbb{I}_{\mathbf{N}}^\mathrm{R} | \mathbb{F}_\mathrm{BG}]_{\partial V} + [\mathbb{I}_{\mathbf{N}}^\mathrm{R} | \mathbb{F}_\mathrm{scat}]_{\partial V} . \label{alpha}
\end{equation}
For the last term, we can use Eq.~(\ref{Fexp}), which yields:
\begin{equation}
[\mathbb{I}_{\mathbf{N}}^\mathrm{R} | \mathbb{F}_\mathrm{scat}]_{\partial V} = -\sum\limits_n  \frac{[\mathbb{I}_{\mathbf{N}}^\mathrm{R} | \mathbb{F}_n]_{\partial V}\langle \mathbb{F}_n^\mathrm{R} | \Delta\hat{\mathbb{M}} | \mathbb{F}_\mathrm{BG} \rangle_V}{k-k_n} . \label{scat}
\end{equation}
It should be noted that $\mathbb{F}_n$ exhibits outgoing boundary conditions, so that it can be decomposed into outgoing fields $\mathbb{O}_\mathbf{N}(\mathbf{r};k_n)$ on the surface $\partial V$. As a remark, the analytic continuation of $\mathbb{F}_n$, which is used in Eq.~(\ref{Norm}), is defined as $
\mathbb{F}_n(\mathbf{r};k) = \sum_{\mathbf{N}}\alpha_{n,\mathbf{N}}\mathbb{O}_{\mathbf{N}}(\mathbf{r};k)$,
with the expansion coefficients $\alpha_{n,\mathbf{N}}$ given by $\alpha_{n,\mathbf{N}} = [\mathbb{I}^\mathrm{R}_\mathbf{N}(k_n)|\mathbb{F}_n]_{\partial V}$~\cite{Weiss2017a,Muljarov2016b,Muljarov2018a}, so that $\mathbb{F}_n(\mathbf{r};k_n) = \mathbb{F}_n(\mathbf{r})$.


Let us now consider a background field $\mathbb{F}_\mathrm{BG}$ that is generated by an incoming field $\mathbb{I}_{\mathbf{N}'}$ and calculate the residue of Eq.~(\ref{alpha}) at $k_n$:
\begin{equation}
\begin{split}
&\mathcal{R}_{n,\mathbf{N}\mathbf{N}'} = \underset{k=k_n}{\operatorname{Res}} \alpha_{\mathbf{N}}^\mathrm{out}(k)=\underset{k=k_n}{\operatorname{Res}} [\mathbb{I}_{\mathbf{N}}^\mathrm{R} | \mathbb{F}_\mathrm{scat}]_{\partial V} \\
&\hspace{0.5cm}= - [\mathbb{I}_{\mathbf{N}}^\mathrm{R} (k_n)| \mathbb{F}_n]_{\partial V} \langle \mathbb{F}_n^\mathrm{R} | \Delta\hat{\mathbb{M}}(k_n) | \mathbb{F}_\mathrm{BG}(k_n) \rangle_V. \label{Help0}
\end{split}
\end{equation}
In order to derive a more explicit expression for the volume integral in the second line, we take Eq.~(\ref{ResDef}) for the reciprocal conjugate resonant state $\mathbb{F}_n^\mathrm{R}$ using $\hat{\mathbb{M}}^\mathrm{R}=\hat{\mathbb{M}}$, multiply it from the left with $\mathbb{F}_\mathrm{BG}(k_n)$, and integrate over a finite volume $V$: $\langle \mathbb{F}_\mathrm{BG}(k_n) |\hat{\mathbb{M}}(k_n)|\mathbb{F}_n^\mathrm{R}\rangle_V = 0$. Using Eq.~(\ref{Background}), we subtract a zero in the form of $
\langle \mathbb{F}_n^\mathrm{R} |\hat{\mathbb{M}}_\mathrm{BG}(k_n)|\mathbb{F}_\mathrm{BG}(k_n)\rangle_V=0$,
which yields due to $\hat{\mathbb{P}}^\mathrm{T} = \hat{\mathbb{P}}$ that
\begin{align}
0=&\langle \mathbb{F}_\mathrm{BG}(k_n) |\hat{\mathbb{M}}(k_n)|\mathbb{F}_n^\mathrm{R}\rangle_V -\langle \mathbb{F}_n^\mathrm{R} |\hat{\mathbb{M}}_\mathrm{BG}(k_n)|\mathbb{F}_\mathrm{BG}(k_n)\rangle_V\nonumber\\
=& \langle \mathbb{F}_n^\mathrm{R} |\Delta\hat{\mathbb{M}}(k_n)| \mathbb{F}_\mathrm{BG}(k_n)\rangle_V \nonumber\\
&-\langle \mathbb{F}_\mathrm{BG}(k_n) |\hat{\mathbb{D}}|\mathbb{F}_n^\mathrm{R}\rangle_V +\langle \mathbb{F}_n^\mathrm{R} |\hat{\mathbb{D}}|\mathbb{F}_\mathrm{BG}(k_n)\rangle_V. \label{Help1}
\end{align}
With the help of Eq.~(\ref{Bilinear3}), we are then able to convert the volume integral in Eq.~(\ref{Help0}) into a surface integral:
\begin{equation}
\langle \mathbb{F}_n^\mathrm{R} |\Delta\hat{\mathbb{M}}(k_n)|\mathbb{F}_\mathrm{BG}(k_n)\rangle_V=[\mathbb{F}_n^\mathrm{R}|\mathbb{F}_\mathrm{BG}(k_n)]_{\partial V}. \label{Inter1}
\end{equation}

On the surface $\partial V$, the field $\mathbb{F}_n^\mathrm{R}$ exhibits outgoing boundary conditions, i.e., it can be constructed as a superposition of outgoing waves $\mathbb{O}^\mathrm{R}_{\mathbf{N}}(\mathbf{r};k_n)$. Owing to this outgoing nature of $\mathbb{F}_n^\mathrm{R}$ and the orthogonality Eq.~(\ref{Orth2}), we obtain that any outgoing field as part of the background field $\mathbb{F}_\mathrm{BG}$ on the right hand side of Eq.~(\ref{Inter1}) does not contribute to the volume integral on the left hand side, so that we can replace $\mathbb{F}_\mathrm{BG}$ by the incoming field. Particularly, by selecting the incoming field to be basis functions $\mathbb{I}_{\mathbf{N}'}(r;k_n)$, we derive that the residues in Eq.~(\ref{Sexp}) are given by
\begin{equation}
\mathcal{R}_{n,\mathbf{N}\mathbf{N}'} = -[\mathbb{I}_{\mathbf{N}}^\mathrm{R}(k_n) | \mathbb{F}_n]_{\partial V}[\mathbb{F}_n^\mathrm{R} |\mathbb{I}_{\mathbf{N}'} (k_n)]_{\partial V}. \label{Residue}
\end{equation}
This is the main result of this work, which allows for calculating the pole contribution in the scattering matrix solely from the resonant field distribution $\mathbb{F}_n$ and the wavenumber $k_n$. Hence, by determining additionally the background term $\mathcal{S}_\mathrm{BG}$, e.g., using a fit or analytical considerations, the fields at any point in space outside the scatterer can be calculated from Eqs.~(\ref{ArbFieldExp}),~(\ref{Smatrix}),~(\ref{Sexp}), and~(\ref{Residue}), while the internal fields are given by Eq.~(\ref{Fexp}). Note that the background term can be reduced to the scattering matrix of homogeneous and isotropic space for some highly symmetric geometries, as it is assumed in the examples of~\cite{Alpeggiani2017a}, but this is not necessarily the case. 

As written above in section~\ref{RS}, the pole expansion of the Green's dyadic in Eq.~(\ref{GreenExp}) should be restricted to a minimal convex volume enclosing the scatterer. Consequently, equation~(\ref{Residue}) is evaluated on the surface of this minimal volume.

\section{Planar periodic systems}\label{PSsec}

As test systems, we consider planar periodic systems with a scattering material distribution that is periodic in the $xy$ plane and bound in the $z$ direction. More specifically, $\hat{\mathbb{M}}(\mathbf{r}+\mathbf{R};k)= \hat{\mathbb{M}}(\mathbf{r};k)$ for any translation vectors of the form $\mathbf{R}=(n_1P_1)\mathbf{a}_1+(n_2P_2)\mathbf{a}_2$ with $P_1$ and $P_2$ being the periods in the directions defined by the unit vectors $\mathbf{a}_1$ and $\mathbf{a}_2$, respectively, and $n_1,n_2\in\mathbb{Z}$. Without the loss of generality, we assume that $\mathbf{a}_1$ and $\mathbf{a}_2$ are normal to the $z$ direction.

The minimal convex volume spans over one unit cell in the periodic directions with a plane on top and bottom that touches the scatterer (see Fig.~\ref{Fig2}b). The orthogonal basis is given by s- and p-polarized plane waves, where `s' specifies linearly polarized light with the electric field being perpendicular to the incidence plane (the plane defined by the incident $\mathbf{k}$ vector and the $z$ axis), while `p' denotes an electric field that is parallel to the incidence plane. The role of the quantum numbers $\mathbf{N}$ is taken by the two polarizations and the reciprocal lattice vectors $\mathbf{G}=(n_1 2\pi/P_1)\mathbf{b}_1+(n_2 2\pi/P_2)\mathbf{b}_2$ with $\mathbf{a}_\alpha\cdot\mathbf{b}_\beta =\delta_{\alpha,\beta}$. Owing to Bloch's theorem, the fields $\mathbb{F}$ as solutions of the scattering at the periodic system can be constructed as a product of a phase factor $\exp(i\mathbf{k}_{||}\cdot\mathbf{r})$ and vector components that are periodic functions with the same periodicity as $\hat{\mathbb{M}}$. The in-plane momentum $\mathbf{k}_{||}$ is preserved apart from Umklapp processes throughout the whole system. Solving Maxwell's equations is then reduced to finding the periodic part of the field distributions for periodic sources or incident fields. It is convenient to introduce a Green's dyadic $\hat{\mathbb{G}}_{\mathbf{k}_{||}}$ for each value of $\mathbf{k}_{||}$, which obeys the following constituting equation:
\begin{equation}
\hat{\mathbb{M}}(\mathbf{r};k) \hat{\mathbb{G}}_{\mathbf{k}_{||}}(\mathbf{r};\mathbf{r}';k) =  \sum\limits_\mathbf{R} \mathbb{1} \mathrm{e}^{i\mathbf{k}_{||}\cdot \mathbf{R}} \delta(\mathbf{r}-\mathbf{r}'-\mathbf{R}) . \label{GreenDef}
\end{equation}
Thus, it is sufficient to consider only one unit cell for further analysis. For instance, any volume of integration can be restricted to span over one unit cell in the $xy$ plane. If a surface integral contains all boundaries of the unit cell, it reduces for periodic integrands to an integration over the top and bottom surfaces of one unit cell, since the surface integrals to adjacent unit cells cancel out due to the periodicity. This implies that we can distinguish the channels of the scattering matrix not only by plane wave orders $\mathbf{G}$ and polarization $p$, but also by the surface at which the fields enter or leave the minimal convex volume (see Fig.~\ref{Fig2}b).

Let us define $\mathbf{K}\equiv \mathbf{k}_{||} + \mathbf{G}$. Then, the orthonormal basis in the top (t) and bottom (b) half space is given by supervectors $\mathbb{F}_{p,\mathbf{K},\pm}^\mathrm{t/b}(\mathbf{r};k)$ with all possible $\mathbf{K}$ for the given $\mathbf{k}_{||}$ and the electric and magnetic fields
\begin{align}
\mathbf{E}_{p,\mathbf{K},\pm}^\mathrm{t/b}(\mathbf{r};k)&= N_{\mathbf{K}}^\mathrm{t/b}(k)\hat{\mathbf{E}}_{p,\mathbf{K},\pm}^\mathrm{t/b}(k) \psi_{\mathbf{K},\pm}^\mathrm{t/b}(\mathbf{r};k), \label{E1}\\ \mathbf{H}_{p,\mathbf{K},\pm}^\mathrm{t/b}(\mathbf{r};k)&= \frac{N_{\mathbf{K}}^\mathrm{t/b}(k)}{Z^\mathrm{t/b}(k)}\hat{\mathbf{H}}_{p,\mathbf{K},\pm}^\mathrm{t/b}(k)\psi_{\mathbf{K},\pm}^\mathrm{t/b}(\mathbf{r};k) , \label{H1}
\end{align}  
where $N_{\mathbf{K}}^\mathrm{t/b}$ is a normalization constant, $\hat{\mathbf{E}}_{p,\mathbf{K},\pm}^\mathrm{t/b}$ and $\hat{\mathbf{H}}_{p,\mathbf{K},\pm}^\mathrm{t/b}$ are unit polarization vectors, $p$ denotes either s or p polarization, $Z^\mathrm{t/b}(k)=\sqrt{\mu^\mathrm{t/b}(k)/\varepsilon^\mathrm{t/b}(k)}$ is the impedance, and
\begin{equation}
\psi_{\mathbf{K},\pm}^\mathrm{t/b}(\mathbf{r};k)=\frac{1}{\sqrt{S_\mathrm{u}}}\mathrm{e}^{i\mathbf{K}\cdot\mathbf{r}\pm i\kappa^\mathrm{t/b}_\mathbf{K}(k) (z-z^\mathrm{t/b})}. \label{Psi}
\end{equation}
Here, $z^\mathrm{t/b}$ gives the positions of the $xy$ planes as the top and bottom surfaces of the minimal convex volume, $S_\mathrm{u}$ is the top/bottom area of one unit cell, and 
\begin{equation}
\kappa^\mathrm{t/b}_\mathbf{K}(k) =  \sqrt{ (k^\mathrm{t/b})^2 - \mathbf{K}^2}, \label{Kz}
\end{equation}
where $k^\mathrm{t/b} = n^\mathrm{t/b}(k)k$ with refractive index $n^\mathrm{t/b}(k) = \sqrt{\varepsilon^\mathrm{t/b}(k)\mu^\mathrm{t/b}(k)}$. For lossless $n^\mathrm{t/b}$ and real wavenumbers $k$, either $\kappa^\mathrm{t/b}_\mathbf{K}$ or $i\kappa^\mathrm{t/b}_\mathbf{K}$ is purely real. In the case of the former, the channel associated with that plane wave is called open, since the field can propagate in $z$ direction from and to the far-field region. In the case of the latter, the channel is closed, and the fields grow or decay exponentially. We select the sign for the square root in Eq.~(\ref{Kz}) such that the fields are either forward-propagating or decaying in positive $z$ direction. For complex $k$, the sign is chosen to match the sign of the same channel on the real axis. If the channel is open on the real axis, the real part of $\kappa^\mathrm{t/b}_\mathbf{K}$ must specify forward propagation in positive $z$ direction. Otherwise, the field must decay in positive $z$ direction. The sign `$\pm$' in Eqs.~(\ref{E1}) to~(\ref{Psi}) thus means forward propagation or decay for `$+$' and backward propagation or decay for `$-$'. In the same manner, it is possible to define incoming and outgoing channels for lossy materials in the top and bottom half spaces by making an analytic continuation from real to complex $n^\mathrm{t/b}$.

The s-polarized unit vectors can be written in the top and bottom half spaces as
\begin{align}
\hat{\mathbf{E}}_{\mathrm{s},\mathbf{K},\pm}^\mathrm{t/b}(k) &= \frac{1}{|\mathbf{K}|}\left(\!\begin{array}{c} -K_y \\ K_x \\ 0 \end{array}\!\right) , \label{TEpolE}\\ 
\hat{\mathbf{H}}_{\mathrm{s},\mathbf{K},\pm}^\mathrm{t/b}(k) &= \frac{1}{k^\mathrm{t/b} |\mathbf{K}|} \left[\!\begin{array}{c} \mp K_x \kappa_\mathbf{K}^\mathrm{t/b}(k) \\ \mp K_y \kappa_\mathbf{K}^\mathrm{t/b}(k) \\ \mathbf{K}^2 \end{array}\!\right] , \label{TEpolH}
\end{align}
The p-polarized unit vectors are given by 
\begin{align}
\hat{\mathbf{E}}_{\mathrm{p},\mathbf{K},\pm}^\mathrm{t/b}(k) &= \mp \hat{\mathbf{H}}_{\mathrm{s},\mathbf{K},\pm}^\mathrm{t/b}(k), \label{TMpolE}\\
\hat{\mathbf{H}}_{\mathrm{p},\mathbf{K},\pm}^\mathrm{t/b}(k)&= \pm \hat{\mathbf{E}}_{\mathrm{s},\mathbf{K},\pm}^\mathrm{t/b}(k). \label{TMpolH}
\end{align}
The normalization constant is
\begin{equation}
N_{\mathbf{K}}^\mathrm{t/b}(k) = \sqrt{\frac{iZ^\mathrm{t/b}(k)k^\mathrm{t/b}}{2\kappa_\mathbf{K}^\mathrm{t/b}(k)}}. \label{NormPW}
\end{equation}

We define the outgoing and incoming channels in the top and bottom half spaces as
\begin{align}
\mathbb{O}_{p,\mathbf{K}}^\mathrm{t}(\mathbf{r};k)&=\mathbb{F}_{p,\mathbf{K},-}^\mathrm{t}(\mathbf{r};k), & \mathbb{O}_{p,\mathbf{K}}^\mathrm{b}(\mathbf{r};k)&=\mathbb{F}_{p,\mathbf{K},+}^\mathrm{b}(\mathbf{r};k), \\
\mathbb{I}_{p,\mathbf{K}}^\mathrm{t}(\mathbf{r};k)&=\mathbb{F}_{p,\mathbf{K},+}^\mathrm{t}(\mathbf{r};k), & \mathbb{I}_{p,\mathbf{K}}^\mathrm{b}(\mathbf{r};k)&=\mathbb{F}_{p,\mathbf{K},-}^\mathrm{b}(\mathbf{r};k).
\end{align}
The reciprocal conjugate is given by
\begin{align}
\left(\mathbb{O}_{p,\mathbf{K}}^\mathrm{t/b}\right)^\mathrm{R} &= \mathbb{O}_{p,-\mathbf{K}}^\mathrm{t/b}, & \left(\mathbb{I}_{p,\mathbf{K}}^\mathrm{t/b}\right)^\mathrm{R} &= \mathbb{I}_{p,-\mathbf{K}}^\mathrm{t/b}, \label{RecPairPW}
\end{align}
which warrants together with Eq.~(\ref{NormPW}) the orthonormality relation given by Eqs.~(\ref{Orth1}) and~(\ref{Orth2}). For details, see Appendix~\ref{OrthPW}. If $\mathbb{F}_n$ is a solution of Eq.~(\ref{ResDef}) for a fixed $\mathbf{k}_{||}$, the reciprocal conjugate resonant state $\mathbb{F}_n^\mathrm{R}$ is a solution of Eq.~(\ref{ResDef}) for $-\mathbf{k}_{||}$ at the same wavenumber $k_n$.

Combining Refs.~\cite{Weiss2017a,Muljarov2018a} (see Appendix~\ref{NormPWsys}), the normalization of the resonant states of the planar periodic systems can be summarized as
\begin{equation}
I_n+\sum\limits_{\mathbf{K}} \frac{1}{2}\left(S_{n,\mathbf{K}}^\mathrm{t} + S_{n,\mathbf{K}}^\mathrm{b}  \right)= 1 , \label{NormPWrec1}
\end{equation}
with 
\begin{align}
I_n &= \langle \mathbb{F}_n^\mathrm{R}|\left.\frac{\partial k\mathbb{P}}{\partial k}\right|_{k_n} |\mathbb{F}_n\rangle_{V_\mathrm{N}} , \label{NormPWrec2}\\
S_{n,\mathbf{K}}^\mathrm{t/b} &=  \left(\beta^\mathrm{t/b}_{n,\mathrm{s},\mathbf{K}}-\beta^\mathrm{t/b}_{n,\mathrm{p},\mathbf{K}}\right)\left.\frac{\partial n^\mathrm{t/b}_{p,\mathbf{K}}}{\partial k}\right|_{k_n}, \label{NormPWrec3}
\end{align}
where 
\begin{equation}
\beta_{n,p,\mathbf{K}}^\mathrm{t/b} = \alpha^\mathrm{t/b}_{n,p,\mathbf{K}}\left(\alpha^\mathrm{t/b}_{n,p,\mathbf{K}}\right)^\mathrm{R}\mathrm{e}^{2i\kappa_\mathbf{K}^\mathrm{t/b}(k_n)\Delta z^\mathrm{t/b}_\mathrm{N}},
\end{equation}
and
\begin{align}
n^\mathrm{t/b}_{\mathrm{s},\mathbf{K}}(k) &= \ln \frac{\kappa_\mathbf{K}^\mathrm{t/b}(k)}{Z^\mathrm{t/b}(k)k^\mathrm{t/b}}, \label{nterms0}\\ n^\mathrm{t/b}_{\mathrm{p},\mathbf{K}}(k) &= \ln \frac{Z^\mathrm{t/b}(k)\kappa_\mathbf{K}^\mathrm{t/b}(k)}{k^\mathrm{t/b}} . \label{nterms}
\end{align}
Here, $\alpha^\mathrm{t/b}_{n,p,\mathbf{K}} = [\mathbb{I}_{p,-\mathbf{K}}^\mathrm{t/b}(k_n)|\mathbb{F}_n ]_{T/B}$ gives the plane wave expansion of the resonant state at the top and bottom surface $T$ and $B$, respectively, enclosing the minimal convex volume $V$. The reciprocal counterpart is $(\alpha^\mathrm{t/b}_{n,p,\mathbf{K}})^\mathrm{R} = [\mathbb{I}_{p,\mathbf{K}}^\mathrm{t/b}(k_n)|\mathbb{F}_n^\mathrm{R} ]_{T/B}$. Furthermore, $\Delta z^\mathrm{t/b}_\mathrm{N}\geq 0$ is the distance between the top and bottom planes of the minimal convex volume $V$ and the normalization volume $V_\mathrm{N}$.

\section{Results}\label{Rsec}

As first example, we consider a planar symmetric slab of refractive index $n^\mathrm{s}=2.5$ and thickness $d=50\,$nm that is surrounded by air ($n^\mathrm{t}=n^\mathrm{b}=1$). The transmission $t$ and reflection $r$ through such a system can be calculated analytically~\cite{BohrenHuffman1983a}. At normal incidence ($\mathbf{k}_{||}=0$), we obtain:
\begin{align}
t(k) &= \frac{\tau\mathrm{e}^{in^\mathrm{s}kd}}{1-\rho^2\mathrm{e}^{2i n^\mathrm{s}kd}} , \label{tslab}\\
r(k) &= \frac{\rho(\mathrm{e}^{2i n^\mathrm{s}kd} - 1)}{1-\rho^2\mathrm{e}^{2i n^\mathrm{s}kd}} . \label{rslab}
\end{align}
In this case, $\rho=(1-n^\mathrm{s})/(1+n^\mathrm{s})$ and $\tau = 4n^\mathrm{s}/(1+n^\mathrm{s})^2$. Equations~(\ref{tslab}) and~(\ref{rslab}) can be summarized in a two-dimensional scattering matrix:
\begin{equation}
\mathcal{S}(k) = \frac{1}{1-\rho^2\mathrm{e}^{2i n^\mathrm{s}kd}} \left[\begin{array}{cc} \rho(\mathrm{e}^{2i n^\mathrm{s}kd} - 1) & \tau\mathrm{e}^{in^\mathrm{s}kd}\\
\tau\mathrm{e}^{in^\mathrm{s}kd} & \rho(\mathrm{e}^{2i n^\mathrm{s}kd} - 1)\end{array}\right] . \label{Sslab}
\end{equation}
This matrix exhibits poles at
\begin{equation}
k_n = \frac{n\pi + i\ln\rho}{n^\mathrm{s}d} .
\end{equation}
The corresponding residues yield
\begin{equation}
\mathcal{R}_n = \frac{2i}{[1-(n^\mathrm{s})^2]d}\left[\begin{array}{cc} 1 & (-1)^{n+1} \\ (-1)^{n+1} & 1\end{array}\right] , \label{ResSlab}
\end{equation}
while the background scattering matrix is given by
\begin{equation}
\mathcal{S}_\mathrm{BG} = \frac{2i}{[1-(n^\mathrm{s})^2]d}\sum\limits_n\frac{1}{k_n}\mathbb{1}, \label{SBGslab}
\end{equation}
where $\mathbb{1}$ is a $2\times2$ unit matrix.

\begin{figure}[htbp]
\begin{center}
\includegraphics[width = \linewidth]{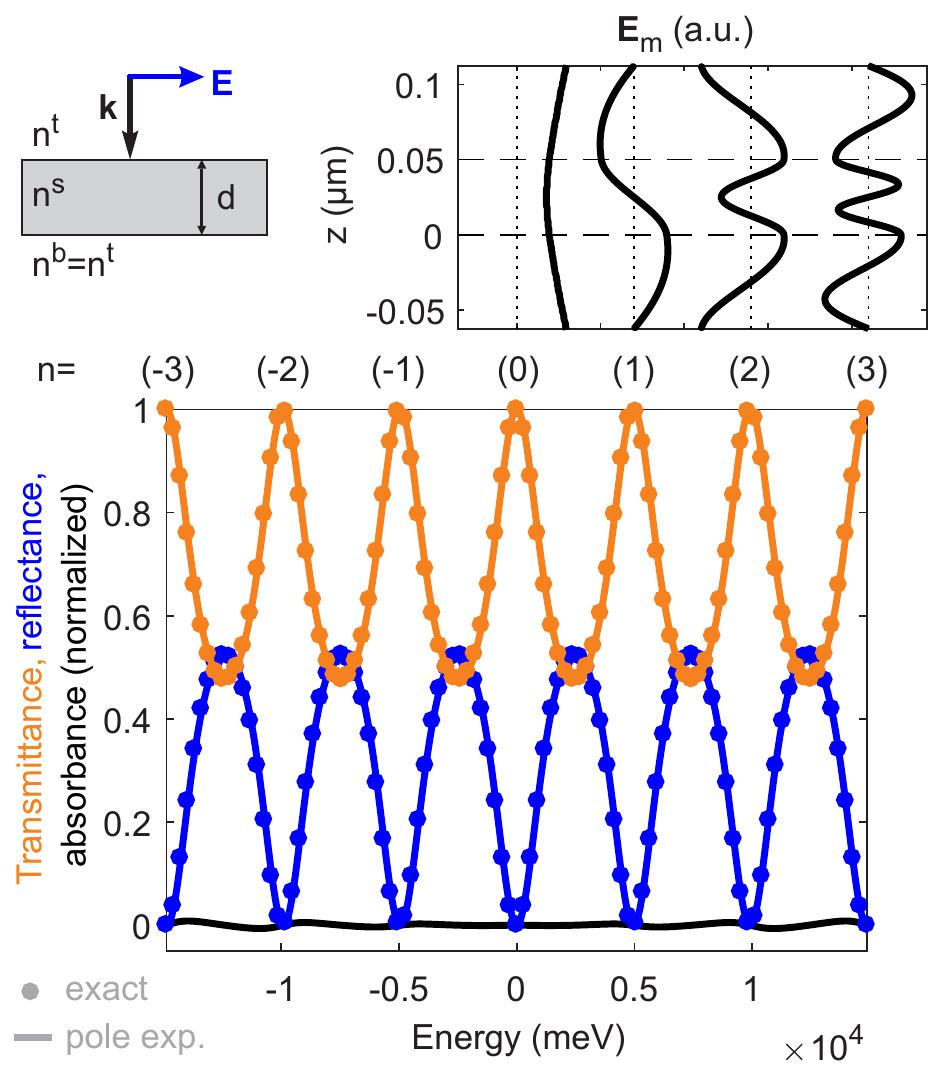}
\caption{(color online) Test system 1: Normalized transmittance (orange), reflectance (blue) and absorbance (black) of a planar symmetric slab of refractive index $2.5$ and thickness $50\,$nm in air at normal incidence (schematic at the top left). The results have been calculated analytically (dots) and based on the pole expansion using $301$ resonant states (lines). The electric field distributions (real part) of the resonant states with index $n=0,1,2,3$ are displayed at the top right.}\label{Fig3}
\end{center}
\end{figure}

Figure~\ref{Fig3} displays the transmittance (orange), reflectance (blue), and absorbance (black) calculated from Eq.~(\ref{Sslab}) (dots) as well as the pole expansion of the scattering matrix (lines). For the pole expansion, we have considered $301$ resonant states symmetrically distributed around $k=0$. The residues have been calculated from the correctly normalized resonant states, which exhibit a perfect agreement with the analytical values given by Eq.~(\ref{ResSlab}). For the background scattering matrix, we have used Eq.~(\ref{SBGslab}) with the sum truncated to the finite number of the $301$ resonant states. Evidently, the pole expansion exhibits a good agreement with the analytical results in the given range, with a growing deviation for larger energies that can be particularly seen in the nonzero absorbance. This deviation becomes smaller when using more resonant states as basis.


\begin{figure}[htpb]
\begin{center}
\includegraphics[width=\linewidth]{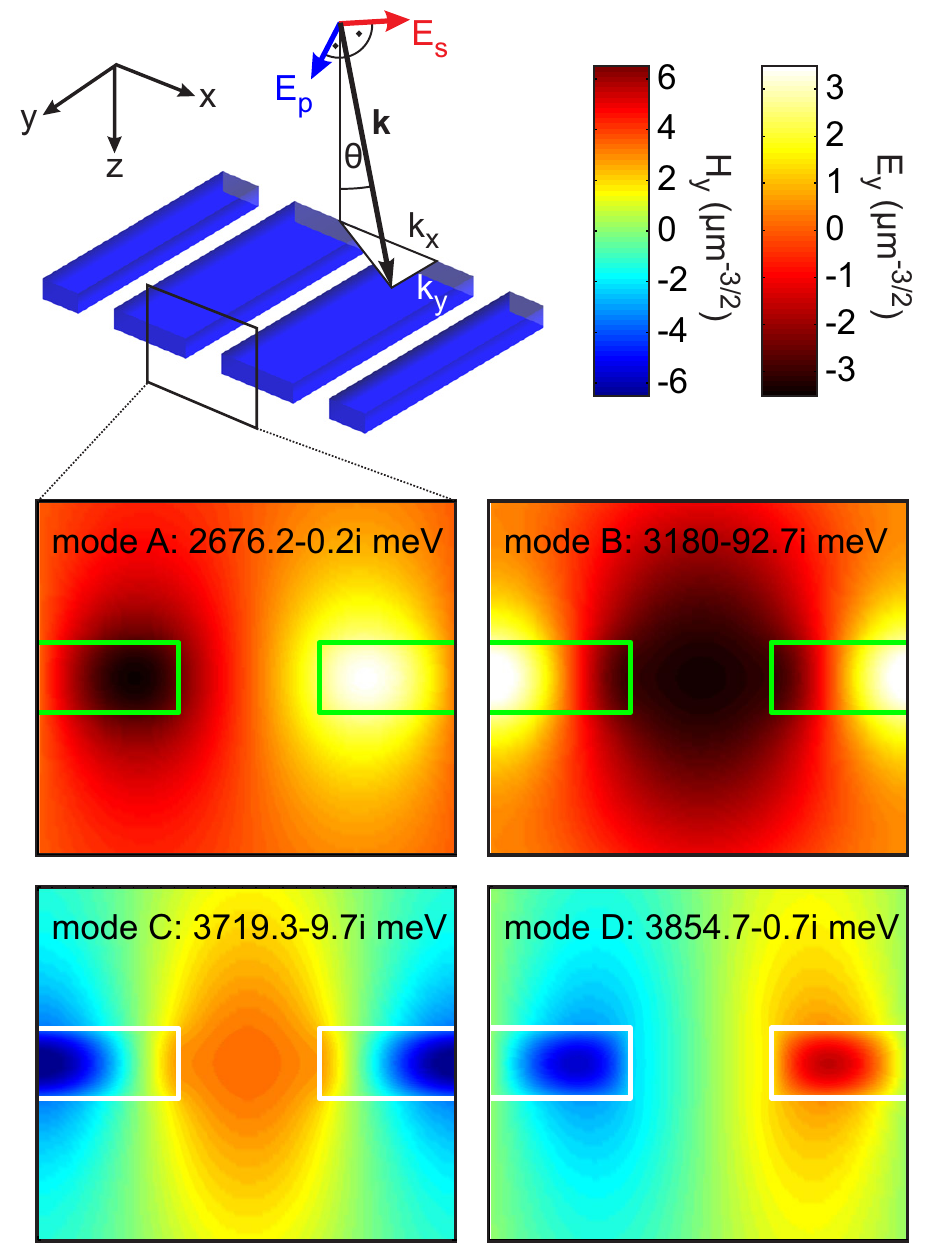}  
\caption{(color online) Test system 2: We consider a dielectric one-dimensional periodic grating with period $300\,$nm and thickness $50\,$nm with high-index regions of refractive index $2.5$ and width $200\,$nm embedded in air. For in-plane wavevector components $k_x=k_y=0.2\,$\textmu$\mathrm{m}^{-1}$, the system exhibits two transverse-electric quasi-guided modes at eigenenergies of $E_\mathrm{A}=2676.2-0.2i\,$meV and $E_\mathrm{B}=3180.0-92.7i\,$meV, respectively, and two transverse-magnetic quasi-guided modes at eigenenergies of $E_\mathrm{C}=3719.3-9.7i\,$meV and $E_\mathrm{D}=3854.7-0.7i\,$meV, respectively. The field plots at the bottom display the real parts of the $y$ components of the normalized electric (modes A and B) and magnetic (modes C and D) resonant field distribution of these modes in one unit cell. The material interfaces are indicated by green and white lines, respectively.} \label{Fig4}
\end{center}
\end{figure}

The second test system is a dielectric one-dimensional periodic grating (see Fig.~\ref{Fig4}) with period $300\,$nm that consists of a high-index material of refractive index $2.5$ with width $200\,$nm and height $50\,$nm embedded in air~\cite{Gippius2010a}. Without the loss of generality, we consider p-polarized incidence for in-plane wavevector components $k_x=k_y=0.2\,$\textmu$\mathrm{m}^{-1}$. In this case, the far-field spectra exhibit four resonant states in the region between $2500\,$meV and $4000\,$meV that are quasi-guided modes~\cite{Tikhodeev2002a}. The corresponding poles are located at $2676.2-0.2i\,$meV (mode A), $3180.0-92.7i\,$meV (mode B), $3719.3-9.7i\,$meV (mode C), and $3854.7-0.7i\,$meV (mode D). The resonant electric and magnetic field of the transverse-electric and transverse-magnetic quasiguided modes, respectively, can be seen in Fig.~\ref{Fig4}. Of course, the system has an infinite number of resonant states as well as cut contributions. However, for the given energy range and polarization, all other resonant states and cuts are far enough away in the complex $k$ plane, so that their impact can be reduced to a slowly varying background.

\begin{figure}[htpb]
\begin{center}
\includegraphics[width=\linewidth]{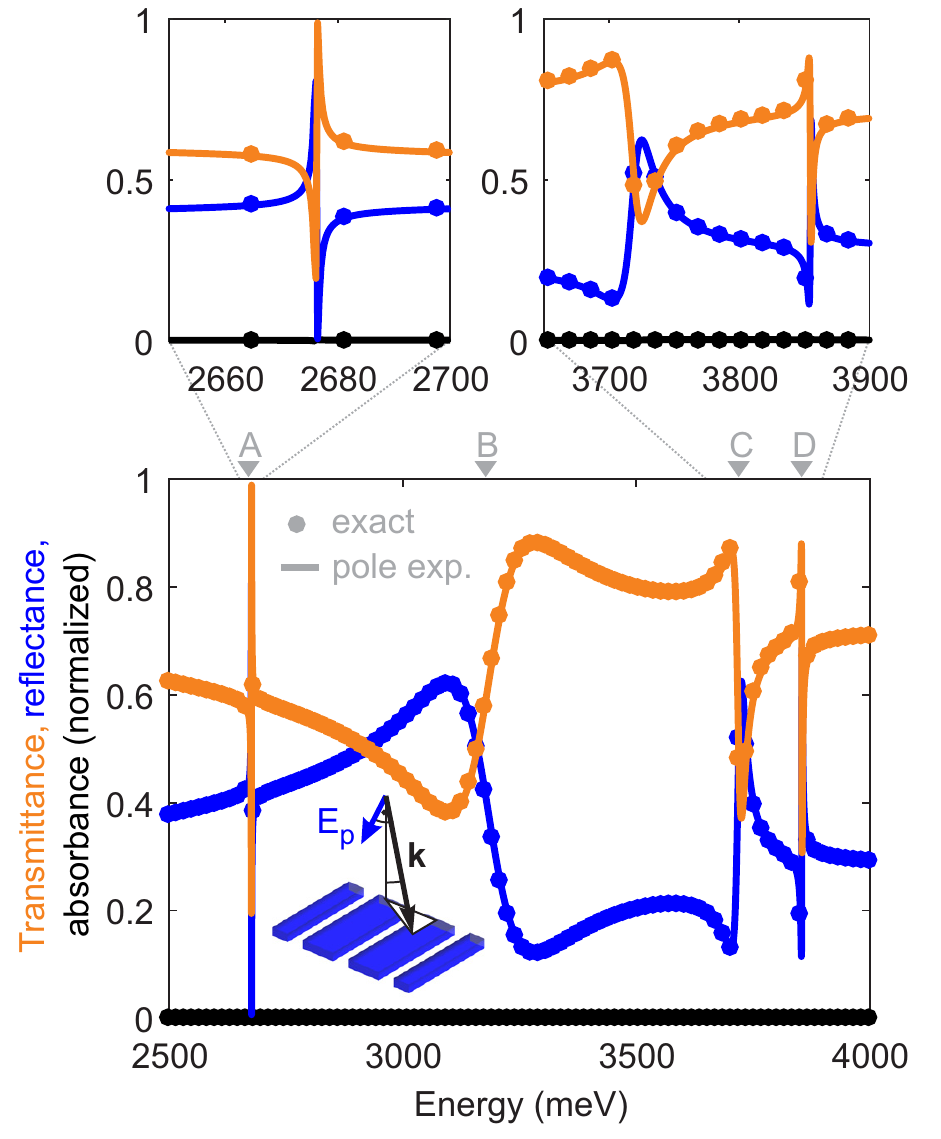}  
\caption {(color online) Transmittance (orange), reflectance (blue), and absorbance (black) of test system 2, derived by full numerical calculations (dots) and the pole expansion of the scattering matrix (solid lines) for p-polarized incidence with $k_x=k_y=0.2\,$\textmu$\mathrm{m}^{-1}$. The gray triangles on top indicate the resonance energy of the four quasi-guided modes in that energy range (see Fig.~\ref{Fig4}). Note that the pole expansion provides an accurate description even around the narrow modes A and D (magnification in top panels), while the full numerical calculations cannot properly resolve the resonant functional behavior on the given equidistant energy grid.} \label{Fig5}
\end{center}
\end{figure}

Figure~\ref{Fig5} displays the far-field spectra of test system~2, with transmittance (orange), reflectance (blue), and absorbance (black). Since the system is nonabsorbing, the absorbance equals zero and is only shown in order to verify that the pole expansion of the scattering matrix fulfills the energy conservation. The filled dots have been calculated by full numerical calculations, whereas the solid lines are derived by using Eqs.~(\ref{Sexp}) and~(\ref{Residue}). The nonresonant background and the influence of other poles and cuts is taken into account by a cubic fit to the spectra calculated at four equidistant energy points in the considered energy range. 

Note that the number of points for the full numerical calculations has been chosen such that the total calculation time equals that for deriving the resonant states and carrying out the quadratic fit for the background. Even though the derivation of the pole expansion is not yet optimized with respect to calculation time -- in contrast to our full numerical approach, it can be seen that the full numerical calculations provide much less details than the pole expansion. This is particularly obvious around the narrow modes A and D (magnification in top panels), where the full numerical calculations cannot resolve the behavior of transmittance and reflectance accurately. Note that the absorbance derived by the pole expansion is less than $0.7\,$\% over the entire energy range. This value can be further reduced by making a higher-order fit or by taking more resonances and some cut contributions into account.

\begin{figure}[htpb]
\begin{center}
\includegraphics[width=\linewidth]{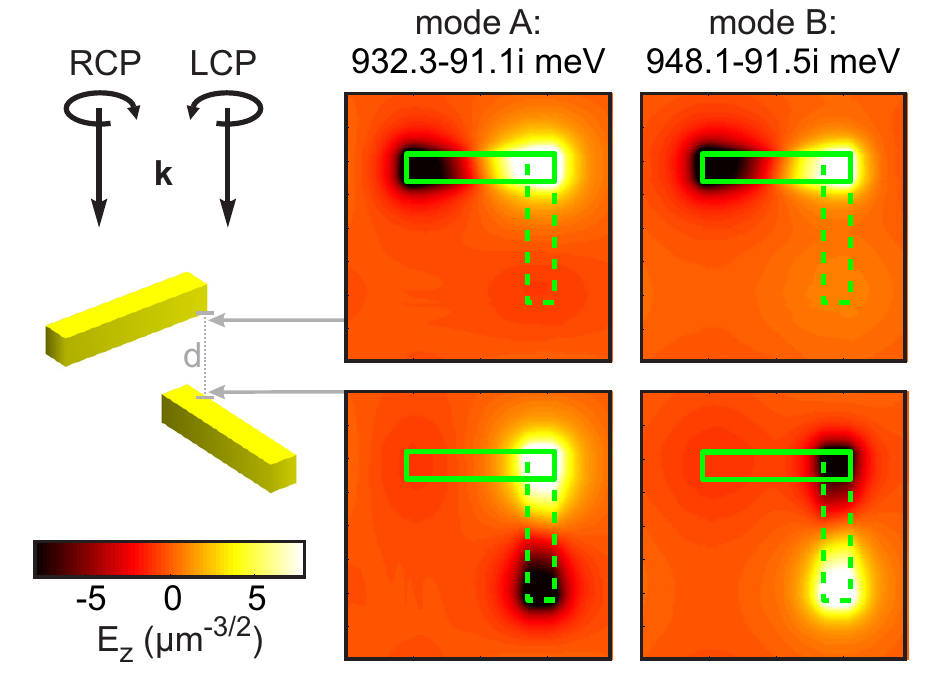}  
\caption {(color online) Test system 3: Chiral arrangement of an array of gold wire pairs with period $400\,$nm in $x$ and $y$ direction. The wires have a width and height of $40\,$nm and are $220\,$nm long. The vertical distance $d$ between the wire pairs is $120\,$nm. The surrounding material has a refractive index of $1.5$. We consider normal incidence, i.e., $k_x=k_y=0$. The two localized plasmon modes that originate in the hybridization of the fundamental dipolar modes of the single wires exhibit eigenenergies of $E_\mathrm{A}=932.3-91.1i\,$meV and $E_\mathrm{B}=948.1-91.5i\,$meV. The field plots on the right display the real parts of the $z$ components of the normalized resonant electric field $10\,$nm below the top wire (top row) and $10\,$nm above the bottom wire (bottom row) in one unit cell. Green solid and dashed lines indicate the location of the upper and lower antenna, respectively.} \label{Fig6}
\end{center}
\end{figure}

Test system 3 consists of a square array of two gold wire antennas of $40\,$nm width and height and $220\,$nm length per unit cell, with period $400\,$nm (see Fig.~\ref{Fig6}). The two antennas are vertically displaced with a distance of $120\,$nm and rotated by $90^\circ$ with respect to each other. The surrounding material has a refractive index of $1.5$. The gold permittivity is described by an analytical model~\cite{Etchegoin2006a}. As shown in Ref.~\cite{Yin2013a}, this system is chiral and exhibits a large circular dichroism as the absorbance difference for left- and right-handed circularly polarized light incidence. Note that we are using here the convention of the point of observer for the definition of the handedness of the light. The circular dichroism originates in the excitation of the bonding and anti-bonding combination of the fundamental plasmon modes in the two wire antennas. Owing to the spatial configuration of the antennas, these modes can be predominately excited by incident light of opposite handedness. Furthermore, they are spectrally shifted. The bonding mode A is located for $\mathbf{k}_{||}=0$  at $932.3-91.1i\,$meV, the anti-bonding mode is at $948.1-91.5i\,$meV. The $z$ component of the corresponding resonant electric field distributions can be seen in Fig.~\ref{Fig6}.

\begin{figure}[htpb]
\begin{center}
\includegraphics[width=\linewidth]{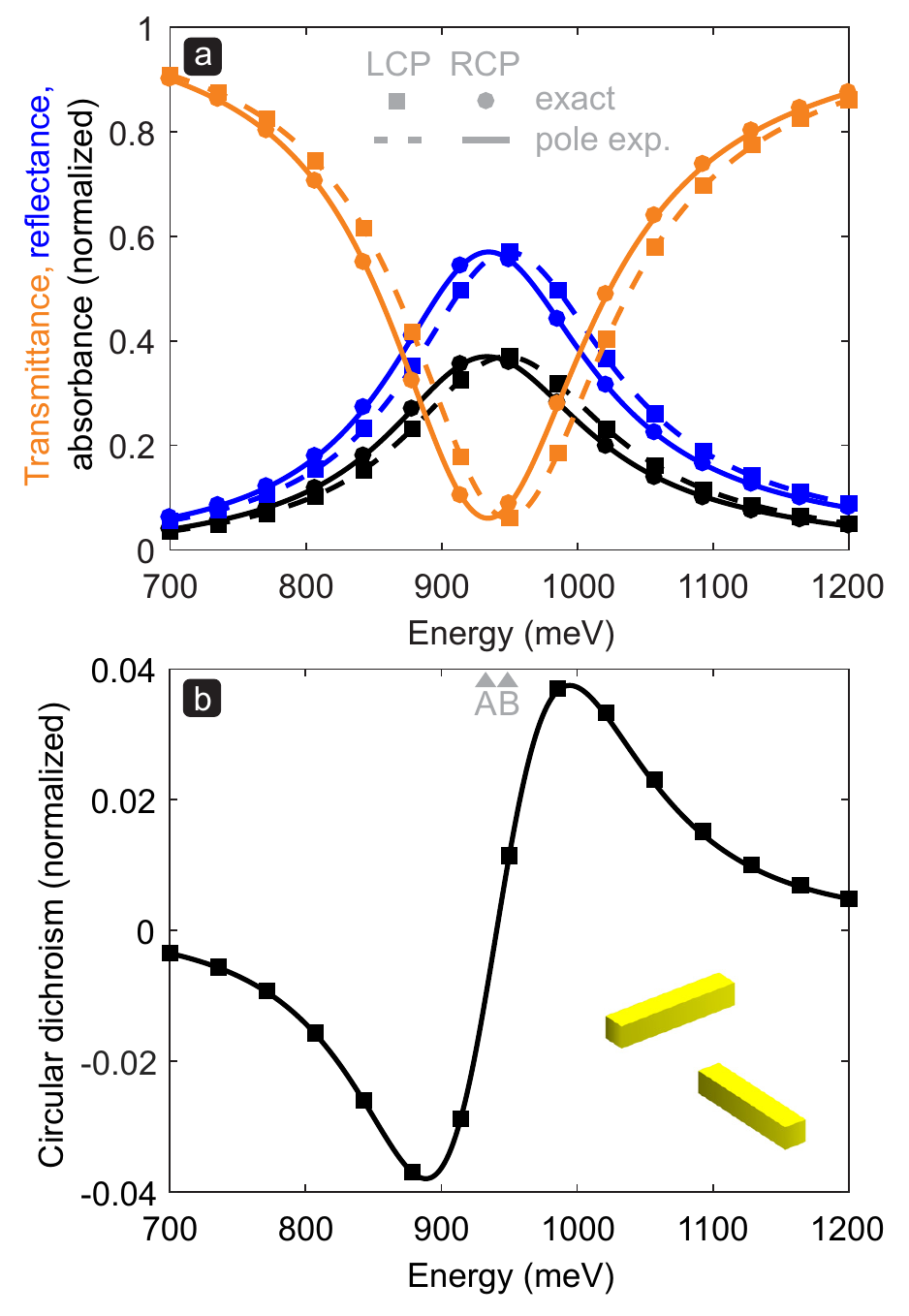}  
\caption {(color online) (a) Transmittance (orange), reflectance (blue), and absorbance (black) of test system 3, derived by full numerical calculations (symbols) and the pole expansion of the scattering matrix (lines) at normal incidence. Solid lines and dots correspond to right-handed circular polarized (RCP) light, whereas dashed lines and squares denote the results for left-handed circularly polarized (LCP) light. (b)~Circular dichroism as the absorbance difference for incident light with left- and right-handed circular polarization, respectively. The gray triangles on top indicate the resonance energy of the two localized plasmon modes in that energy range (see Fig.~\ref{Fig6}).} \label{Fig7}
\end{center}
\end{figure}

While the nonabsorbing test systems 1 and 2 can be treated by the formulation of the pole expansion described in Ref.~\cite{Alpeggiani2017a}, the third test system is beyond its scope due to the large Ohmic losses. Its absorbance reaches values of nearly $40\,$\%, as seen in Fig.~\ref{Fig7}a. The far-field spectra are displayed for left-handed circularly polarized (LCP, dashed lines and squares) and right-handed circularly polarized (RCP, solid lines and dots) plane waves at normal incidence. The pole expansion (solid and dashed lines) agrees well with the full numerical calculation (dots and squares) and predicts the circular dichroism correctly (see Fig.~\ref{Fig7}b).

Note that any contributions to the pole expansion beyond modes A and B have been treated in similar way as in test system~2 by carrying out a quadratic fit to three equidistant energy points in the range between $700\,$ meV and $900\,$meV. Furthermore, the calculation time for deriving the selected number of energy points by the full numerical calculations roughly equals the time for deriving the two modes and the quadratic fit. As in the case of test system 2, it is obvious that the pole expansion provides much more details than the full numerical results within the same calculation time. This is due to the fact that the pole expansion of the scattering matrix in Eq.~(\ref{Sexp}) has an explicit analytic dependence on wavenumber, so that once all relevant resonant states in the wavenumber range of interest are calculated and projected onto the basis functions in free space, the scattering matrix becomes immediately available in the whole wavenumber range.

\section{Conclusion}

We have developed a formulation of the pole expansion of the scattering matrix that allows for deriving the residues of the pole contributions solely from the resonant states. The accuracy and efficiency of the formulation is demonstrated for three test systems, where the pole expansion is compared with exact analytical and numerical results. The pole expansion agrees well with the exact results and provides much more details of the far-field spectra for the same computational time, owing to the analytic wavenumber dependence. This is particularly useful for optimizing certain geometries in order to obtain desired far-field properties. Moreover, the artificial frequency discretization used in conventional frequency-domain solvers is replaced by a natural discretization in terms of the physically meaningfull resonant states, which can provide intuitive insights into the underlying physical mechanisms.

\section*{Acknowledgement}

T. W. acknowledges support from DFG SPP 1838, the VW-Foundation and the MWK Baden-W{\"u}rttemberg. E. A. M. acknowledges support by the EPSRC Grant EP/M020479/1 and RBRF Grant 16-29-03283. 

\appendix

\section{Single particles} \label{Spherical}

In the case of single particles in three-dimensional space, the minimal convex volume has to be chosen as a sphere. On a unit sphere, the vector spherical harmonics $\mathbf{Y}_{lm}$, $\mathbf{\Psi}_{lm}$, and $\mathbf{\Phi}_{lm}$ are forming a complete orthogonal basis~\cite{Barrera1985a}, with
\begin{align}
\mathbf{Y}_{lm} &=\hat{\mathbf{e}}_rY_{lm} , & \mathbf{\Psi}_{lm}&=r\nabla Y_{lm}, & \mathbf{\Phi}_{lm} &= \mathbf{r}\times\nabla Y_{lm},
\end{align}
where $Y_{lm}$ are scalar orthonormal spherical harmonics, $r=|\mathbf{r}|$, and $\hat{\mathbf{e}}_r$ is the unit vector in the radial direction. The vector spherical harmonics obey the orthogonality relations
\begin{align}
\oint\limits_\Omega \mathrm{d}\Omega \,\mathbf{Y}_{l'm'}^*\cdot \mathbf{Y}_{lm}&= \delta_{l,l'}\delta_{m,m'} , \label{VHS1}\\
\oint\limits_\Omega \mathrm{d}\Omega \,\mathbf{\Psi}_{l'm'}^*\cdot \mathbf{\Psi}_{lm} &= l(l+1)\delta_{l,l'}\delta_{m,m'} ,  \label{VHS2}\\
\oint\limits_\Omega \mathrm{d}\Omega \,\mathbf{\Phi}_{l'm'}^*\cdot \mathbf{\Phi}_{lm} &= l(l+1)\delta_{l,l'}\delta_{m,m'} , \label{VHS3}
\end{align}
and
\begin{equation}
\begin{split}
0 &=\oint\limits_\Omega \mathrm{d}\Omega \,\mathbf{Y}_{l'm'}^*\cdot \mathbf{\Psi}_{lm} \\
&=\oint\limits_\Omega \mathrm{d}\Omega \,\mathbf{Y}_{l'm'}^*\cdot \mathbf{\Phi}_{lm}\\
&=\oint\limits_\Omega \mathrm{d}\Omega \,\mathbf{\Psi}_{l'm'}^*\cdot \mathbf{\Phi}_{lm}. \label{VHS4}
\end{split}
\end{equation}
Here, $d\Omega$ is the differential of the solid angle $\Omega$, and the integration is carried out over the whole unit sphere. Furthermore,
\begin{align}
\mathbf{Y}_{lm}^* &= (-1)^m \mathbf{Y}_{l,-m}, \\ \mathbf{\Psi}_{lm}^* &= (-1)^m \mathbf{\Psi}_{l,-m}, \\ \mathbf{\Phi}_{lm}^* &= (-1)^m \mathbf{\Phi}_{l,-m}.
\end{align}

As in the case of plane waves, the spherical-wave solutions of Maxwell's equations have two orthogonal polarizations in homogeneous and isotropic space, labeled transverse-electric (TE) and transverse-magnetic (TM). In this case, transverse-electric and transverse-magnetic means the absence of a radial electric or magnetic field component, respectively. The role of the components of the index vector $\mathbf{N}$ is taken by the two polarizations, the azimuthal quantum number $m$ and the polar number $l$ of the vector spherical harmonics. The basis functions are given by~\cite{Jackson1999a}
\begin{align}
\mathbf{E}_{\mathrm{TE},lm}(\mathbf{r};k) &=  \frac{N_lf_l(kr)}{\sqrt{l(l+1)}} \mathbf{\Phi}_{lm}(\Omega), \label{ETE}\\ 
\mathbf{H}_{\mathrm{TE},lm}(\mathbf{r};k) &=  \frac{-iN_l}{Z^\mathrm{s}k\sqrt{l(l+1)}} \nabla\times f_l(kr)\mathbf{\Phi}_{lm}(\Omega), \label{HTE}\\
\mathbf{E}_{\mathrm{TM},lm}(\mathbf{r};k) &= \frac{N_l}{k\sqrt{l(l+1)}} \nabla\times f_l(kr)\mathbf{\Phi}_{lm}(\Omega), \label{ETM}\\ 
\mathbf{H}_{\mathrm{TM},lm}(\mathbf{r};k) &=  \frac{-iN_lf_l(kr)}{Z^\mathrm{s}\sqrt{l(l+1)}} \mathbf{\Phi}_{lm}(\Omega) ,\label{HTM}
\end{align}
where $f_l$ is the radial dependence, which is given by spherical Bessel and Neumann functions $j_l$ and $n_l$, respectively, or by outgoing and incoming spherical Hankel functions $h_l^{(+)} = j_l+in_l$ and $h_l^{(-)} = j_l-in_l$, respectively. In addition, $N_l$ is the normalization constant that is determined by Eq.~(\ref{Orth1}), and $Z^\mathrm{s}$ is the impedance of the surrounding homogeneous and isotropic space, while $k$ is here the wavenumber in the surrounding medium. Note that the angular dependence of Eqs.~(\ref{ETE}) to~(\ref{HTM}) is solely given by the vector spherical harmonics, because
\begin{equation}
\nabla\times f_l\mathbf{\Phi}_{lm} = -\frac{1}{r}\left[l(l+1)f_l\mathbf{Y}_{lm}+\frac{\partial rf_l}{\partial r} \mathbf{\Psi}_{lm}\right]. \label{VecId1}
\end{equation}

While the orthogonality relations of the vector spherical harmonics given by Eqs.~(\ref{VHS1}) to~(\ref{VHS4}) contain complex conjugation, the complex conjugation is replaced by a reciprocal conjugation in Eqs.~(\ref{Orth1}) and~(\ref{Orth2}). In order to use the orthogonality properties of the vector spherical harmonics, we chose 
\begin{equation}
\mathbb{F}_{p,lm}^\mathrm{R}(\mathbf{r};k) = (-1)^{m}\mathbb{F}_{p,l,-m}(\mathbf{r};k), \label{RecSP}
\end{equation}
where $\mathbb{F}_{p,l,-m}$ is the supervector consisting of the electric and magnetic fields given by Eqs.~(\ref{ETE}) to (\ref{HTM}). Thus, by using Eq.~(\ref{VecId1}) as well as
\begin{align}
\hat{\mathbf{e}}_r\cdot \left(\mathbf{\Phi}^*_{l'm'} \times \mathbf{Y}_{lm}\right) &= 0, \\ 
\hat{\mathbf{e}}_r\cdot \left(\mathbf{\Phi}^*_{l'm'} \times \mathbf{\Psi}_{lm}\right) &= -\mathbf{\Psi}^*_{l'm'} \cdot \mathbf{\Psi}_{lm} ,
\end{align}
we obtain after some algebra that
\begin{equation}
\begin{split}
 &i \oint\limits_\Omega \mathrm{d}\Omega \,\hat{\mathbf{e}}_r\cdot\left(\tilde{\mathbf{E}}_{p',l'm'}^\mathrm{R}\times \mathbf{H}_{p,lm}\right) = \\
&\hspace{0.5cm}=\frac{N_l^2}{Z^\mathrm{s}} \delta_{l,l'}\delta_{m,m'}\delta_{p,p'}\begin{cases}\frac{\tilde{f}_l}{kr}\frac{\partial rf_l}{\partial r} &\mathrm{for}\,p=\mathrm{TE},\\ \frac{-f_l}{kr}\frac{\partial r\tilde{f}_l}{\partial r} &\mathrm{for}\,p=\mathrm{TM},\end{cases}
\end{split}
\end{equation}
where the tilde indicates that the first field might have a different radial dependence than the second field. 

Let us now define the outgoing and incoming fields $\mathbb{O}_{p,lm}$ and $\mathbb{I}_{p,lm}$, respectively, such that their radial dependence is given by the outgoing and incoming spherical Hankel functions. Thus, the anti-symmetric form of Eq.~(\ref{Bilinear2}) results in Eq.~(\ref{Orth2}) for identical Hankel functions. For pairs of fields with incoming and outgoing Hankel functions, respectively, it is proportional to the Wronskian 
\begin{equation}
W[h_l^{(+)},h_l^{(-)}]=h_l^{(+)}\frac{\partial h_l^{(-)}}{\partial x} - h_l^{(-)}\frac{\partial h_l^{(+)}}{\partial x}=\frac{-2i}{x^2},
\end{equation}
with $x$ denoting the argument of the spherical Hankel functions, which yields Eq.~(\ref{Orth1}) for a spherical volume of radius $R$, provided that 
\begin{equation}
N_l = \sqrt{\frac{-Z^\mathrm{s}}{R^2W[h_l^{(+)}(kR),h_l^{(-)}(kR)]}} = k\sqrt{\frac{Z^\mathrm{s}}{2i}}.
\end{equation}

We note that the background field $\mathbb{F}_\mathrm{BG}$ that fulfills Eq.~(\ref{Background}) is understood to be without singularities. Thus, the radial dependence of the background field must be given by a spherical Bessel function $j_l$. This means that the superposition of incoming and outgoing fields $\mathbb{I}_{p,lm}$ and $\mathbb{O}_{p,lm}$ removes the diverging contribution of the spherical Hankel function.

For single particles, the reciprocal conjugate $\mathbb{F}_n^\mathrm{R}$ can be deduced from $\mathbb{F}_n$ by going from azimuthal order $m$ to $-m$. The analytical normalization of the resonant states in Eq.~(\ref{Norm}) can be simplified by using~\cite{Muljarov2018a}
\begin{equation}
\mathbb{F}'_n = \frac{1}{k_n} (\mathbf{r}\cdot\nabla)\mathbb{F}_n.
\end{equation}

\section{Orthogonality of plane waves} \label{OrthPW}

We show here how to derive the reciprocal conjugate basis functions for the plane waves defined in Eqs.~(\ref{E1}) and~(\ref{H1}) that fulfill the orthonormality relations given by Eqs.~(\ref{Orth1}) and~(\ref{Orth2}). First, note that
\begin{equation}
\int\limits_{S_\mathrm{u}}\mathrm{d}x\mathrm{d}y\,  \psi_{-\mathbf{K}',\pm}^\mathrm{t/b}(\mathbf{r};k) \psi_{\mathbf{K},\pm}^\mathrm{t/b}(\mathbf{r};k) = \delta_{\mathbf{K},\mathbf{K}'},
\end{equation}
from which we identify that the reciprocal conjugate basis functions are solutions for the opposite in-plane momentum $-\mathbf{K}$.

Next, we consider cross products of electric and magnetic components for reciprocal conjugate pairs of basis functions for the same polarization (s or p):
\begin{align}
\hat{\mathbf{E}}_{\mathrm{s},-\mathbf{K},d'}^\mathrm{t/b} \times \hat{\mathbf{H}}_{\mathrm{s},\mathbf{K},d}^\mathrm{t/b} &= \frac{1}{k^\mathrm{t/b}} \left(\begin{array}{c}  -K_x  \\ -K_y  \\ -s_d \kappa_\mathbf{K}^\mathrm{t/b}\end{array}\right) ,\\
\hat{\mathbf{E}}_{\mathrm{p},-\mathbf{K},d'}^\mathrm{t/b} \times \hat{\mathbf{H}}_{\mathrm{p},\mathbf{K},d}^\mathrm{t/b} &=  \frac{1}{k^\mathrm{t/b}} \left(\begin{array}{c}  s_d s_{d'} K_x  \\ s_d s_{d'} K_y  \\ -s_d  \kappa^\mathrm{t/b}_\mathbf{K}\end{array}\right) .
\end{align}
Here, $d,d'=\pm$, with $s_\pm=\pm1$, which specifies the direction of propagation or decay. Note that for different polarizations and the same $\mathbf{K}$, the cross products of the electric and magnetic components vanishes. Therefore, 
\begin{equation}
\hat{\mathbf{n}}_z \cdot \left(\hat{\mathbf{E}}_{p',-\mathbf{K},d'}^\mathrm{t/b}\times \hat{\mathbf{H}}_{p,\mathbf{K},d}^\mathrm{t/b} \right)= \pm\frac{\kappa^\mathrm{t/b}_\mathbf{K}}{k^\mathrm{t/b}} \delta_{p,p'}. \label{nz}
\end{equation}
In the case that the magnetic field corresponds to an outgoing (or decaying) field, the sign on the right hand side of Eq.~(\ref{nz}) is negative. Otherwise, the sign is positive. Hence, for reciprocal pairs of basis functions with identical outgoing or incoming boundary conditions, the antisymmetric form of Eq.~(\ref{Bilinear2}) results in a cancelation of the integrands, which provides Eq.~(\ref{Orth2}). For pairs with different boundary conditions, the sign change in Eq.~(\ref{nz}) prohibits the cancelation, resulting in
\begin{equation}
[\mathbb{I}_{p',-\mathbf{K}'}^\mathrm{t/b}|\mathbb{O}_{p,\mathbf{K}}^\mathrm{t/b}]_{T/B} = N_{\mathbf{K}}^\mathrm{t/b}N_{-\mathbf{K}}^\mathrm{t/b}\frac{2\kappa_\mathbf{K}^\mathrm{t/b}}{iZ^\mathrm{t/b}k^\mathrm{t/b}} \delta_{p',p}\delta_{\mathbf{K}',\mathbf{K}} .
\end{equation}
Since none of the quantities in the fraction on the right hand side depends on the sign of $\mathbf{K}$, we can define $N_{-\mathbf{K}}^\mathrm{t/b}=N_{\mathbf{K}}^\mathrm{t/b}$. Hence, equations~(\ref{NormPW}) and~(\ref{RecPairPW}) provide the orthonormality condition required in Eq.~(\ref{Orth1}).

\section{Normalization of resonant states in planar periodic systems} \label{NormPWsys}

Based on the plane wave expansion of the resonant states in the exterior, it is possible to derive the surface contribution in the normalization condition given by Eq.~(\ref{Norm}). The analytical continuation of the fields in the top and bottom half space is given by 
\begin{align}
\mathbb{F}_n^\mathrm{t/b}(\mathbf{r};k) &= \sum\limits_{p,\mathbf{K}}\alpha_{n,p,\mathbf{K}}^\mathrm{t/b}\mathbb{O}_{p,\mathbf{K}}^\mathrm{t/b}(\mathbf{r};k), \\
\left(\mathbb{F}_n^\mathrm{t/b}\right)^\mathrm{R}(\mathbf{r};k) &= \sum\limits_{p,\mathbf{K}}\left(\alpha_{n,p,\mathbf{K}}^\mathrm{t/b}\right)^\mathrm{R}\left(\mathbb{O}_{p,\mathbf{K}}^\mathrm{t/b}\right)^\mathrm{R}(\mathbf{r};k),
\end{align}
where the expansion coefficients $\alpha_{n,p,\mathbf{K}}^\mathrm{t/b}$ and $(\alpha_{n,p,\mathbf{K}}^\mathrm{t/b})^\mathrm{R}$ are calculated at the top and bottom surface of the minimal convex volume $V$. Owing to the periodicity, the surface integrals to adjacent unit cells in Eq.~(\ref{Norm}) cancel out. For the remaining surface integrals at the top and bottom of the normalization volume $V_\mathrm{N}$, we obtain:
\begin{equation}
\begin{split}
&\Big[\left(\mathbb{F}_n^\mathrm{t/b}\right)^\mathrm{R}\Big|\left(\mathbb{F}_n^\mathrm{t/b}\right)'\Big]_{T_\mathrm{N}/B_\mathrm{N}} = \\
&\hspace{0.5cm}=\sum\limits_{p,\mathbf{K}} \alpha_{n,p,\mathbf{K}}^\mathrm{t/b} \left(\alpha_{n,p,\mathbf{K}}^\mathrm{t/b}\right)^\mathrm{R} \Big[\left(\mathbb{O}_{p,\mathbf{K}}^\mathrm{t/b}\right)^\mathrm{R}\Big|\left(\mathbb{O}_{p,\mathbf{K}}^\mathrm{t/b}\right)'\Big]_{T_\mathrm{N}/B_\mathrm{N}}.
\end{split}
\end{equation}
The surface integral on the right hand side depends on the polarization, because either the electric or the magnetic polarization vectors, equations~(\ref{TEpolH}) and~(\ref{TMpolE}), depend on $k$, while their counterparts Eqs.~(\ref{TEpolE}) and~(\ref{TMpolH}) vanish after differentiation with respect to $k$. After some algebra, we end up with 
\begin{equation}
\Big[\left(\mathbb{O}_{p,\mathbf{K}}^\mathrm{t/b}\right)^\mathrm{R}\Big|\left(\mathbb{O}_{p,\mathbf{K}}^\mathrm{t/b}\right)'\Big]_{T_\mathrm{N}/B_\mathrm{N}} = \pm\frac{1}{2}\left.\frac{\partial n^\mathrm{t/b}_{p,\mathrm{K}}}{\partial k}\right|_{k_n} \mathrm{e}^{2i\kappa_\mathbf{K}^\mathrm{t/b}\Delta z^\mathrm{t/b}_\mathrm{N}}, 
\end{equation}
where the positive sign has to be taken for s polarization, while the minus sign corresponds to p polarization, and $\Delta z^\mathrm{t/b}_\mathrm{N}$ is the distance between the top and bottom planes of the minimal convex volume $V$ and the normalization volume $V_\mathrm{N}$. This results in the normalization of the resonant states for planar periodic systems given by Eqs.~(\ref{NormPWrec1}) to~(\ref{nterms}). Equating the explicit form of the derivatives of the terms $n^\mathrm{t/b}_{p,\mathrm{K}}$ defined in Eqs.~(\ref{nterms0}) and~(\ref{nterms}) yields
\begin{align}
\begin{split}
\left.\frac{\partial n^\mathrm{t/b}_{\mathrm{s},\mathrm{K}}}{\partial k}\right|_{k_n} =& \left[\frac{|\mathbf{K}|}{\kappa_\mathbf{K}^\mathrm{t/b}(k_n)} \right]^2 \frac{1}{k_n^\mathrm{t/b}} \left.\frac{\partial k^\mathrm{t/b}}{\partial k}\right|_{k_n} \\
&- \frac{1}{Z^\mathrm{t/b}(k_n)}\left.\frac{\partial Z^\mathrm{t/b}}{\partial k}\right|_{k_n} , 
\end{split} \\
\begin{split}
\left.\frac{\partial n^\mathrm{t/b}_{\mathrm{p},\mathrm{K}}}{\partial k}\right|_{k_n} =& \left[\frac{|\mathbf{K}|}{\kappa_\mathbf{K}^\mathrm{t/b}(k_n)} \right]^2 \frac{1}{k_n^\mathrm{t/b}} \left.\frac{\partial k^\mathrm{t/b}}{\partial k}\right|_{k_n} \\
&+ \frac{1}{Z^\mathrm{t/b}(k_n)}\left.\frac{\partial Z^\mathrm{t/b}}{\partial k}\right|_{k_n} .
\end{split}
\end{align}




%

\end{document}